\documentclass[twocolumn,numberedappendix]{openjournal}

\usepackage{graphicx}

\usepackage{amsmath,amssymb,amstext}
\usepackage{amsbsy,amsfonts,amsthm,color,bm}
\usepackage[colorlinks,linkcolor=blue,citecolor=blue,urlcolor=blue ]{hyperref}
\usepackage{orcidlink}

\usepackage[utf8]{inputenc}
\setcitestyle{numbers,square}
\usepackage[caption=false]{subfig}

\def\pderiv(#1/#2){\mathchoice{\partial{#1}\over\partial{#2}}
    {\partial{#1}/\partial{#2}} {\partial{#1}/\partial{#2}}
    {\partial{#1}/\partial{#2}}}
\def\pderivp(#1/#2){\left(\pderiv({#1}/{#2})\right)}

\def\tderiv(#1/#2){\mathchoice{d{#1}\over d{#2}}{d{#1}/d{#2}}
                   {d{#1}/d{#2}}{d{#1}/d{#2}}}
\def\tderivp(#1/#2){\left(\tderiv({#1}/{#2})\right)}

\def\hover#1#2{\setbox0\hbox to 0pt{\hss$\scriptscriptstyle\rightharpoonup$}%
               #1\kern0.1ex\raise #2ex\box0\kern-0.1ex}
\def\hover#1#2{\mbox{\boldmath$#1$}}

\def\half{{\textstyle\frac12}}

\def\phie{\phi_E}

\def\mnras{MNRAS}

\def\apj{ApJ}
\def\apjl{ApJL}
\def\aap{A\&A}
\def\araa{ARAA}

\def\nat{Nature}

\def\prd{Phys Rev D}

\let\citet=\citep

\begin{document}

\title{What are the parities of photon-ring images near a black hole?}

\author{Ashish Kumar Meena\orcidlink{0000-0002-7876-4321}$^\star$}
\thanks{$^\star$\href{mailto:ashishmeena766@gmail.com}{ashishmeena766@gmail.com}}
\affiliation{Physics Department, Ben-Gurion University of the Negev, PO Box 653, Be’er-Sheva 8410501, Israel}

\author{Prasenjit Saha\orcidlink{0000-0003-0136-2153}$^\dagger$}
\thanks{$^\dagger$\href{mailto:psaha@physik.uzh.ch}{psaha@physik.uzh.ch}}
\affiliation{Physik-Institut, University of Zurich, Winterthurerstrasse 190, CH-8057 Zurich, Switzerland}

\begin{abstract}

Light that grazes a black-hole event horizon can loop around one or
more times before escaping again, resulting for distance observers in
an infinite sequence of ever fainter and more delayed images near the
black hole shadow.  In the case of the M87 and Sgr~A$^*$ back holes,
the first of these so-called photon-ring images have now been
observed.  A question then arises: are such images minima, maxima, or
saddle-points in the sense of Fermat's principle in gravitational
lensing?  or more briefly, the title question above.  In the theory of
lensing by weak gravitational fields, image parities are readily found
by considering the time-delay surface (also called the Fermat
potential or the arrival-time surface).  In this work, we extend the
notion of the time delay surface to strong gravitational fields and
compute the surface for a Schwarzschild black hole.  The time-delay
surface is the difference of two wavefronts, one travelling forward
from the source and one travelling backwards from the observer.  Image
parities are read off from the topography of the surface, exactly as
in the weak-field regime, but the surface itself is more complicated.
Of the images, furthest from the black hole and similar to the
weak-field limit, are a minimum and a saddle point.  The strong field
repeats the pattern, corresponding to light taking one or more loops
around the back hole.  In between, there are steeply-rising walls in the time-delay
surface, which can be interpreted as maxima and saddle points that are
infinitely delayed and not observable --- these correspond to light
rays taking a U-turn around the black hole.

\end{abstract}

\maketitle

\section{Introduction}
\label{sec:intro}

One of the tests of Einstein's theory of gravity is the deflection of 
light rays (also known as gravitational lensing) by matter 
distributions~\citep{1911AnP...340..898E, 1916AnP...354..769E, 
1936Sci....84..506E}. 
It was first recognised in the observation of stars behind the Sun 
during the famous 1919 eclipse~\citep{1920RSPTA.220..291D} and, later, 
in the radio observation of a distant quasar lying behind a foreground 
galaxy and forming multiple images~\citep{1979Natur.279..381W}. 
Since then, light deflection has been observed from individual stars 
in our Galaxy to distant galaxy clusters~\citep{1986BAAS...18R1014L, 
1987A&A...172L..14S} and became an integral part of the study of the 
Universe~\citep{1992ARA&A..30..311B, 2010CQGra..27w3001B}. 

In general, a weak-field approximation~(where the spacetime can be 
decomposed into a background and a small perturbation on this background 
created by the lens~\citep{2010arXiv1010.3416P}) is sufficient to 
study the conventional gravitational lensing from a star, galaxy, or 
galaxy cluster and explain all observed properties of the lensed 
images~\citep{1992grle.book.....S}.
However, a weak field approximation breaks down very close to a neutron 
star or a black hole, where light rays experience very strong 
gravitational fields.
The existence of such objects (neutron stars and black holes) has been 
confirmed by different methods~(for example, using x-ray 
binaries~\citep{2016A&A...587A..61C}; gravitational wave 
observations~\citep{2016PhRvD..93l2003A, 2018PhRvL.121p1101A}; 
astrometric microlensing~\citep{2022ApJ...933...83S})
and even black-hole shadows have been imaged for
the supermassive black holes at the 
centre of the nearby galaxy M87~\citep{2019ApJ...875L...1E, 
2019ApJ...875L...2E, 2019ApJ...875L...3E, 2019ApJ...875L...4E} 
and our own Galaxy~\citep{2022ApJ...930L..12E, 2022ApJ...930L..13E, 
2022ApJ...930L..14E}.

In the strong gravitational field near a black hole, instead of using the 
conventional lens equation derived using a weak-field approximation, 
one needs to solve the geodesic equation to determine the path of 
light rays.
The simplest case to study light propagation in a strong gravitational 
field is lensing by a Schwarzschild black 
hole~\citep{1916SPAW.......189S}~(also see~\citep{1999physics...5030S}), 
a classic topic in the literature.
An analytical solution for the deflection angle near Schwarzschild 
black hole can be derived in terms of elliptic 
integral~\citep{1959RSPSA.249..180D, ohanian-87, 1983mtbh.book.....C}.
\citet{2000PhRvD..62h4003V} and \citet{2001GReGr..33.1535B} obtained 
(approximate) gravitational lens equation applicable in strong 
gravitational field near the Schwarzschild black hole and discussed 
the presence of the infinite sequence of~(increasingly de-magnified) 
lensed images of a background source (also known as 
\textit{relativistic images}).
Using a different formalism,~\citet{2000PhRvD..61f4021F} derived an 
exact lens equation for Schwarzschild black hole.
Going beyond lensing by Schwarzschild black hole, similar analyses 
have also been performed for Kerr(-Newman) and more exotic black 
holes~\citep[e.g.,][]{2002PhRvD..66j3001B, 2003PhRvD..67j3006B, 
2011JMP....52i2502A, 2020PhRvD.101d4031G, 2021PhRvD.103j4063H}.

\begin{figure*}
    \centering
    \includegraphics[width=1\textwidth]{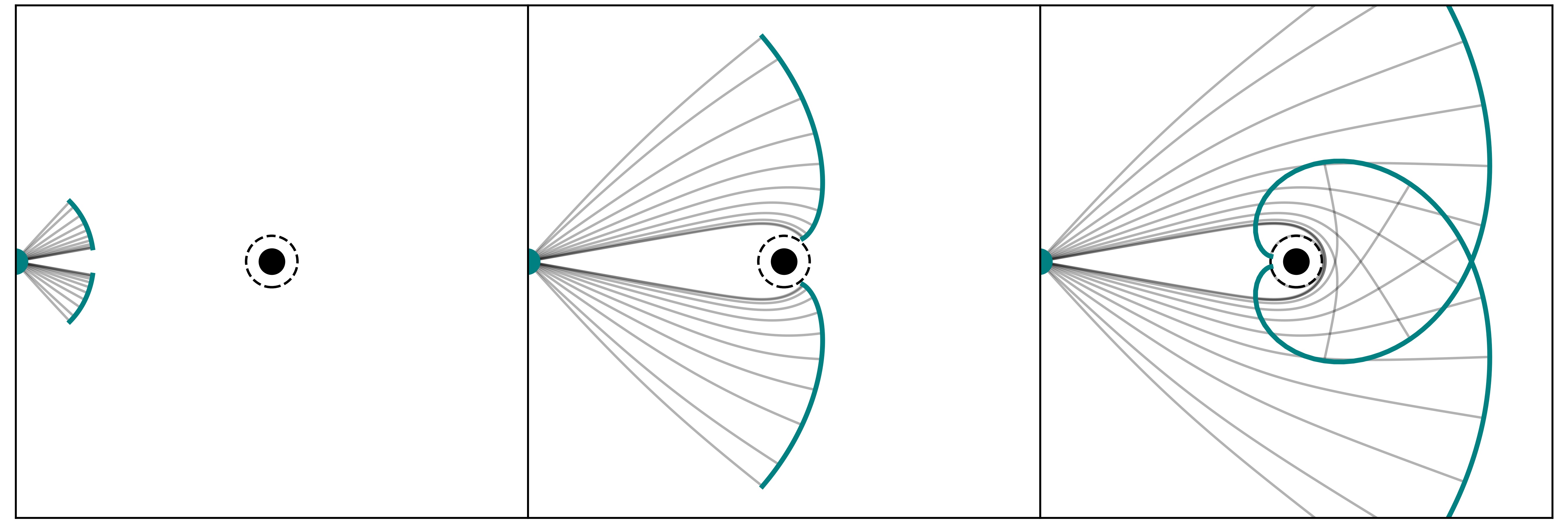}
    \caption{Wavefront propagation near a Schwarzschild 
    black hole lens. The black dot in each panel represents the black 
    hole position. The black dashed circle around the black hole marks 
    the photon sphere~($r=3M$). The green dot represents the source 
    position, and grey curves represent the light rays emitted from 
    the source. The equal time surfaces~(or wavefronts) for these 
    rays are shown by the green curves. The left, middle, and right 
    panels represent the wavefront for three different time values~(in 
    increasing order), respectively. In each panel, the gap in the
    middle of the wavefront corresponds to rays which fall inside
    the black hole.}
    \label{fig:wavefront}
\end{figure*}

A very instructive way to describe gravitational lensing is thinking 
in terms of \textit{wavefronts} emitted from the source instead of 
individual light rays. 
The wavefront method was first used in gravitational lensing to 
estimate the time delay between multiple images for point mass and 
axially symmetric lenses to determine the Hubble 
constant~\citep{1964MNRAS.128..295R, 1964MNRAS.128..307R, 
1966MNRAS.132..101R}. 
For a given lens system (made of source, lens, and observer), a 
wavefront emitted from the source gets deformed and develops 
crossings as it crosses the lens and moves 
forward~\citep[e.g.,][]{2002JMP....43.5578F}.
A pedagogical introduction to wavefronts in gravitational lensing is 
presented in~\citet{Nityananda_1990, 10.1007/BFb0009227}.
The use of the wavefront method in the strong gravitational field was first 
discussed in~\citet{1994ApJ...421...46R} to construct caustic 
structure near a Kerr black hole.
Later, the wavefront method was used in~\citet{2012JMPh....3.1882F,
2014PhRvD..90b3014Y, 2019GReGr..51...32K} to further understand the 
light propagation near Kerr black hole and in other~(more exotic) 
spacetimes~\citep[e.g.,][]{2018GReGr..50....7K}.

In the contemporary literature on lensing in the weak field limit, the
\textit{time delay surface} is a fundamental quantity which can be
used to describe the various properties of the lensed images like
position, magnification, and parity~\citep{1986ApJ...310..568B}.  For
strong fields, there is a general formulation of Fermat's
principle~\citep{1992PhRvD..45.3862N}, but the weak-field time-delay
surface has not been generalised.  There is, however, an elegant
construction using wavefronts~\citep{Nityananda_1990,10.1007/BFb0009227}, which can be
adapted.  In our current work, we use wavefronts to compute time delay
surfaces near the Schwarzschild black hole for axially and non-axially
symmetric cases.  Since the images are essentially extrema points of
the time delay surface, we can infer that even in the strong
gravitational field, the lensed images should be either minima or
maxima, or saddle-point.  However, to determine the exact order in
which these different types of images will appear is not obvious.  In
addition, a Schwarzschild black hole is a singular lens.  This can
lead to tears in the time delay surface similar to the point mass lens
in weak field limit and make the overall geometry of the time delay
surface very complex near the black hole.  Hence, an explicit
construction of time delay surface near the black hole is necessary to
address the above issues.

The current work is organised as follows.
In Section~\ref{sec:nullg}, we briefly discuss the light propagation 
and wavefronts near the Schwarzschild black hole.
In Section~\ref{sec:axis_sym}, we construct the time delay surface 
for axis-symmetric case (i.e., when the source lies on the optical 
axis, a line joining the observer and lens) and discuss the parity of 
the lensed images near the Schwarzschild black hole.
In Section~\ref{sec:off_axis}, similarly, we construct the time delay 
surface and determine the parities for an off-axis source.
Section~\ref{sec:maxima}, discusses the formation of infinitely 
delayed images in between different pairs of observed images.
We conclude our work in Section~\ref{sec:conclusions}.
Throughout this work, we use the natural unit system,~($c=1,\:G=1$), 
unless mentioned otherwise.

\section{Light paths past a Schwarzschild lens}
\label{sec:nullg}

The trajectory of photons passing near the Schwarzschild black hole 
are determined by the following equations, 
\begin{equation}
    \begin{aligned}
    {dr \over dt} &= \left(1-\frac{2M}{r}\right)  \sqrt{1-\frac{b^3}{b-2M}{r-2M \over r^3}}, \\
    {d\phi \over dt} &= \left(1-\frac{2M}{r}\right) {1 \over r^2} \sqrt{{b^3\over b-2M}},
    \label{eq:eq_to_solve}
    \end{aligned}
\end{equation}
where~$r=b$ is the distance of close approach and~$(t,r,\phi)$ are the 
spacetime coordinates in the plane where the light 
travels~(i.e.,~$\theta=\pi/2$).
We refer readers to Appendix~\ref{sec:app1} describing a method to
derive the above equations and Appendix~\ref{sec:app2} for various 
limiting cases of the above equations. 

\begin{figure}
    \centering
    \includegraphics[height=0.46\textwidth,width=0.48\textwidth]{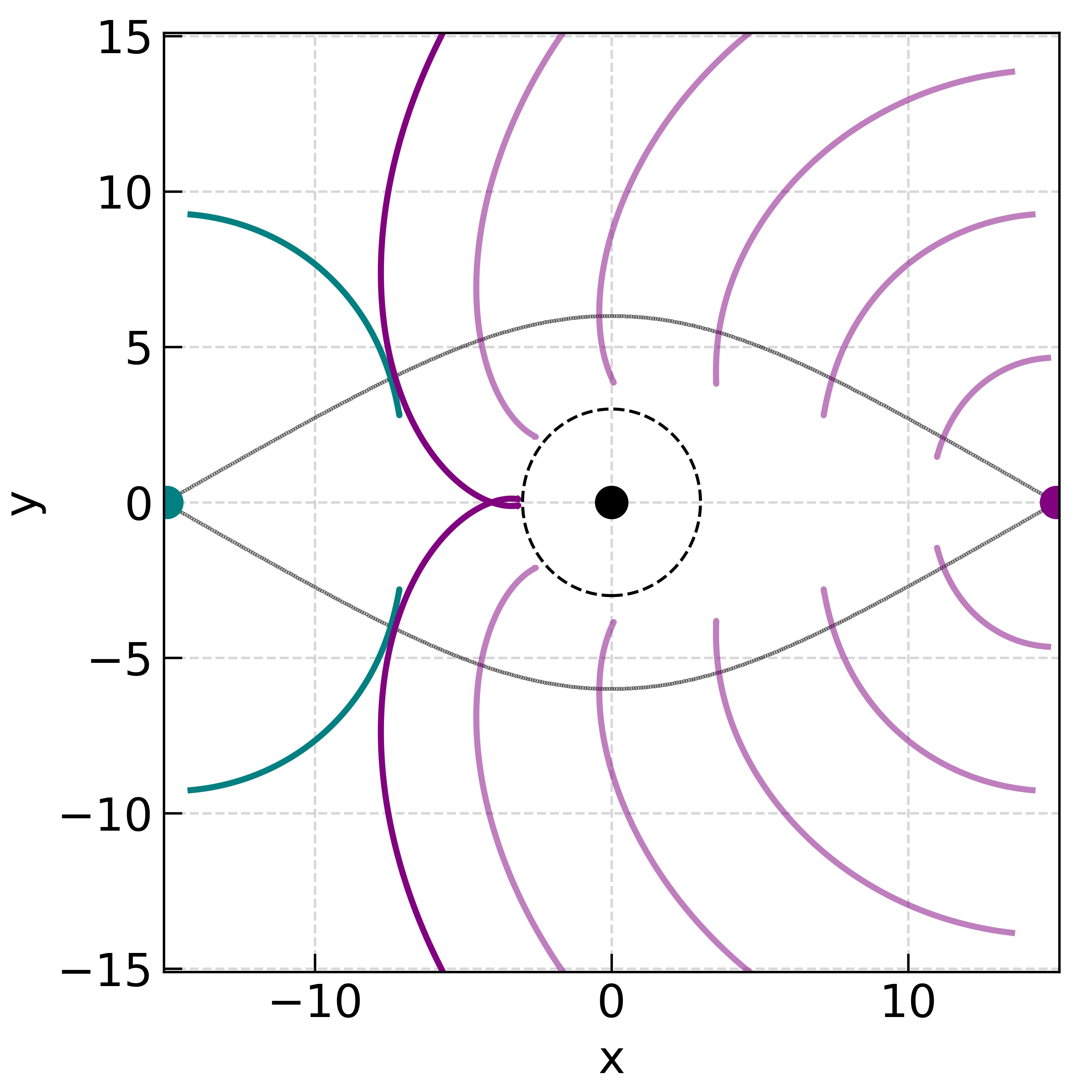}
    \caption{Wavefront propagation near a Schwarzschild black hole in 
    the axially symmetric case. The green, black and purple dots represent 
    the position of the source, lens, and observer, respectively. The green 
    curves represent the wavefront emitted by the source. The purple 
    curves represent a wavefront emitted from the observer at different 
    time instances. The lensed images correspond to points 
    on the purple and green wavefronts where their normal vector agree with 
    each other. These points are essentially the points in the figure 
    where purple and green wavefronts touch each other. The grey curves 
    represent the corresponding rays emitted from the source and reach 
    to the observer.}
    \label{fig:wavefront_touch}
\end{figure}

\begin{figure*}
    \centering
    \includegraphics[width=0.45\textwidth]{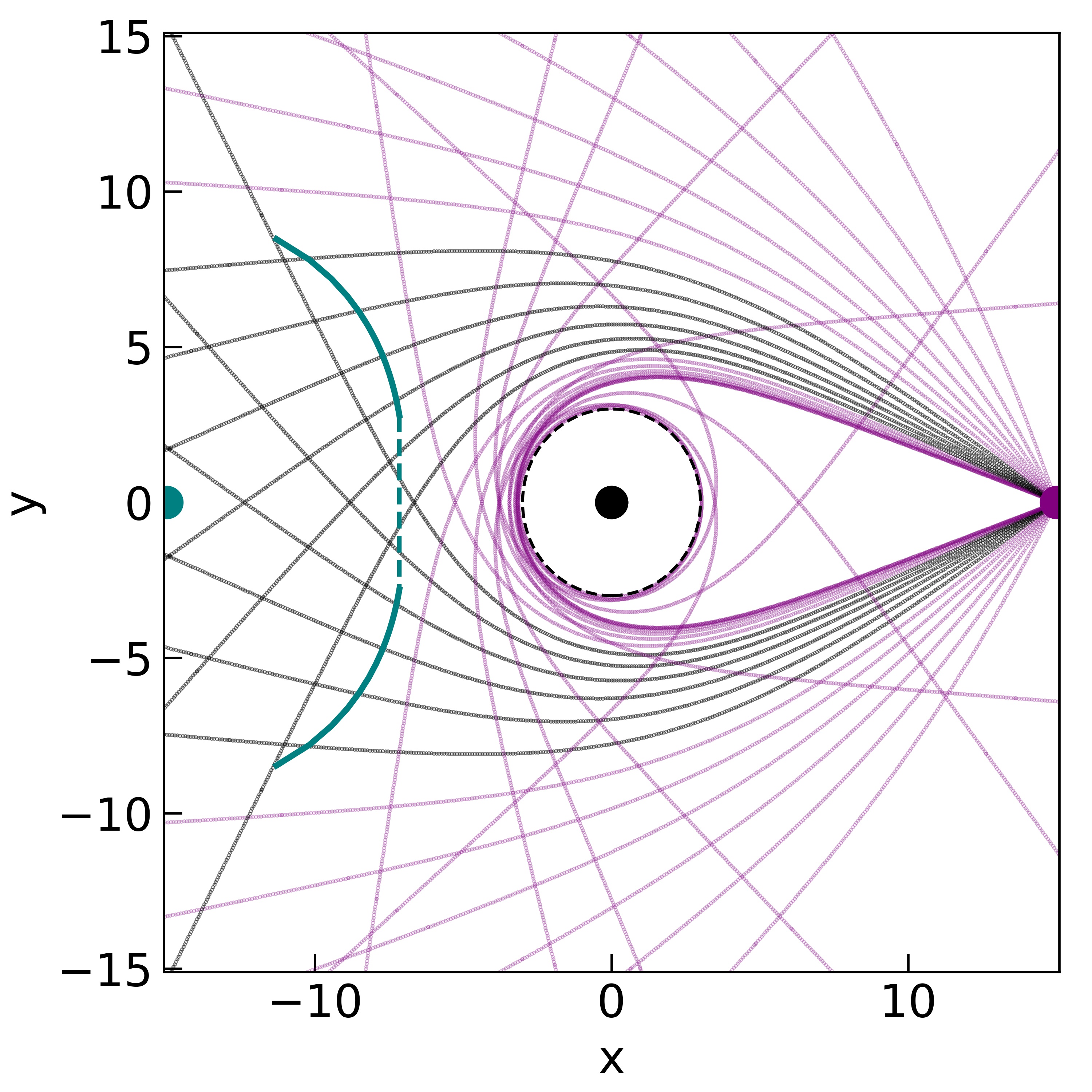}
    \includegraphics[width=0.45\textwidth]{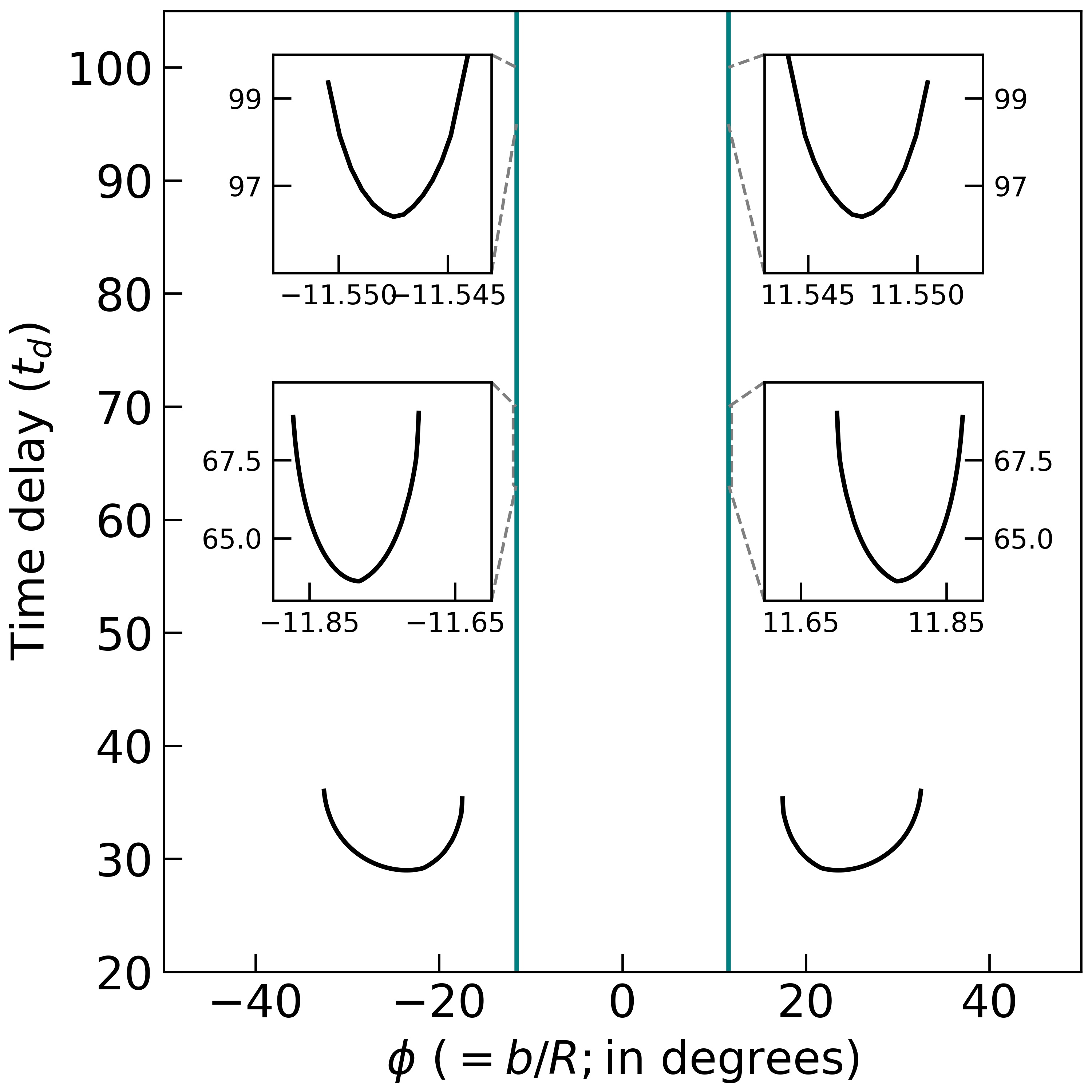}
    \caption{Construction of a time delay surface near a Schwarzschild 
    black hole for the axially symmetric case.
    \textit{Left panel}: The green, black, and purple dots at (-15, 0), 
    (0,0), and (0,15) represent the position of source, Schwarzschild 
    black hole, and observer, respectively. The black dashed 
    line represents the photon sphere~($r=3$) around the black hole. 
    The green curve represents the wavefront emitted by the source. 
    The grey (purple) curves mark the rays emitted from the observer, 
    which (do not) cross the green wavefront. 
    \textit{Right panel}: Time delay~($t_d$; in units of~$GM/c^3$) as 
    a function of angle of closest approach~($\phi$; in degrees) with 
    respect to the observer. The 
    green vertical lines mark the angle corresponding to the photon 
    sphere. The black curves represent the slices of time delay 
    surface near primary, secondary, and tertiary images as we go from 
    small to large time delays. For time delays near the secondary 
    and tertiary images, we show zoom-in plots since the images 
    form very close to the photon sphere.}
    \label{fig:wavefront_td_axis}
\end{figure*}

\begin{figure*}
    \centering
    \includegraphics[width=0.31\textwidth]{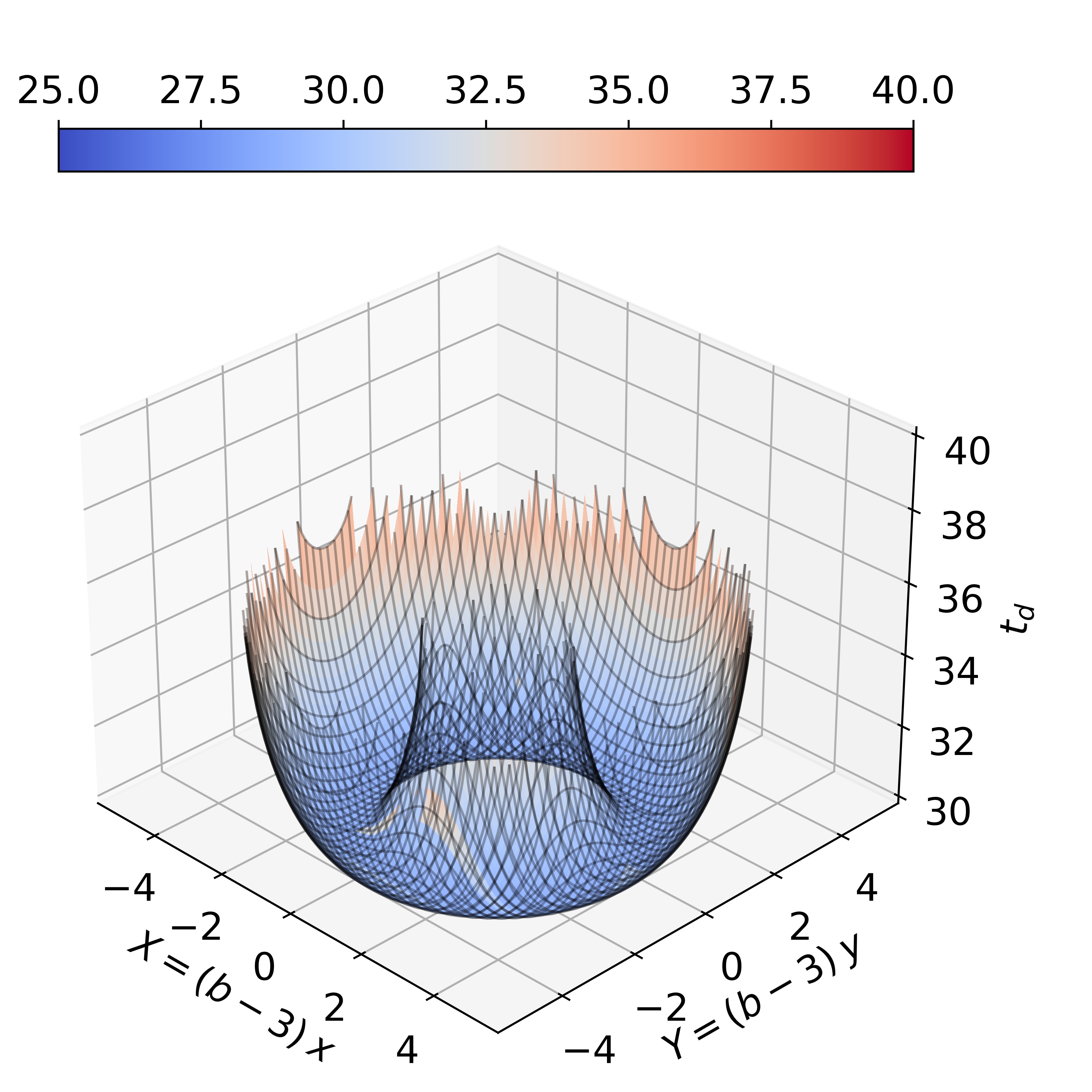}
    \includegraphics[width=0.31\textwidth]{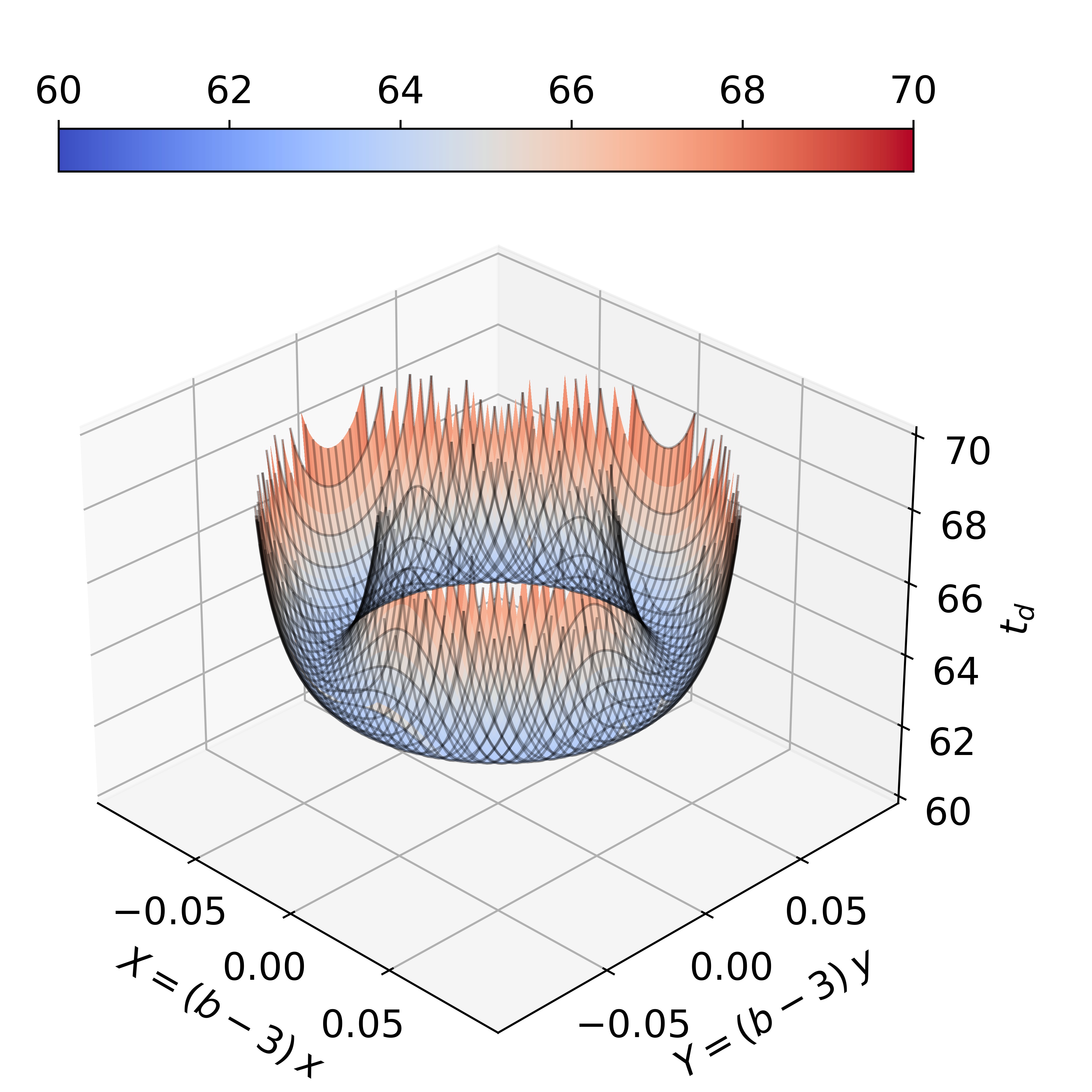}
    \includegraphics[width=0.31\textwidth]{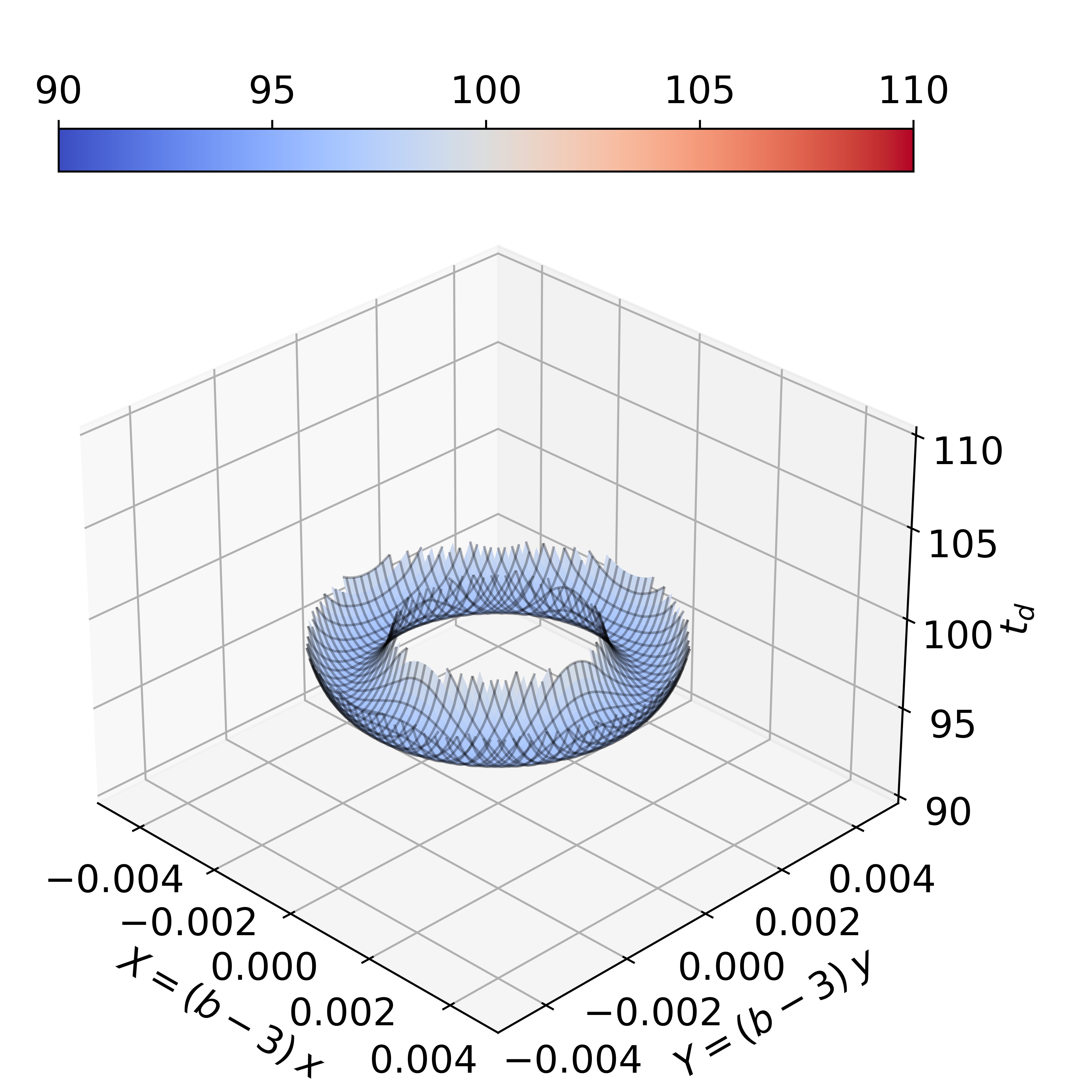}
    \caption{Time delay surface near primary/secondary/tertiary ring 
    in the left/middle/right panel for axially symmetric case. The horizontal
    axes represent the spatial axes of the lens with respect to 
    the observer~(also equivalent to observer sky) and vertical axis represents 
    the time delay values (in units of~$GM/c^3$). In each panel, the
    $X$- and $Y$-axis have been transformed to remove the region inside the photon
    sphere.}
    \label{fig:tds_axis}
\end{figure*}

\begin{figure*}
    \centering
    \includegraphics[width=0.45\textwidth]{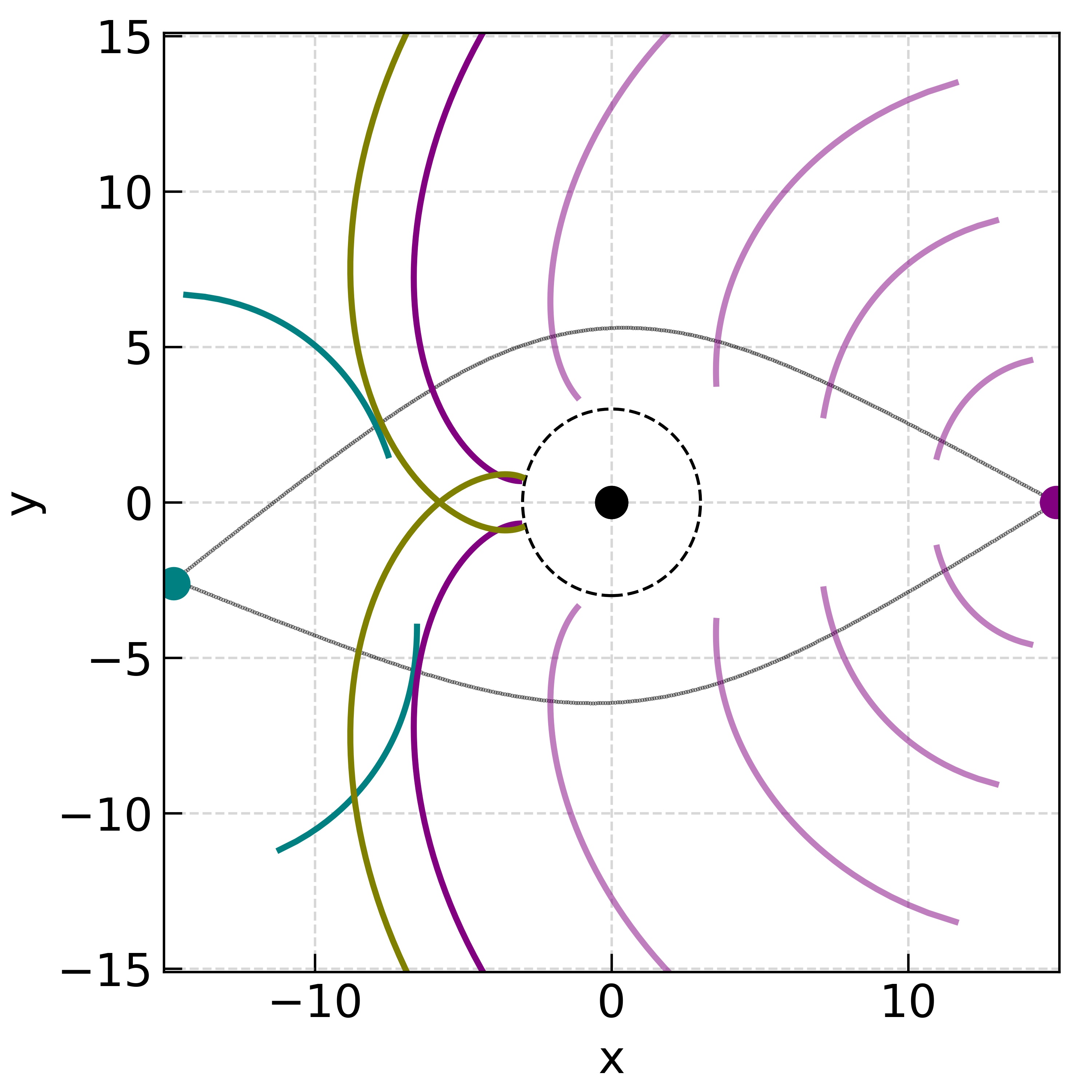}
    \includegraphics[width=0.45\textwidth]{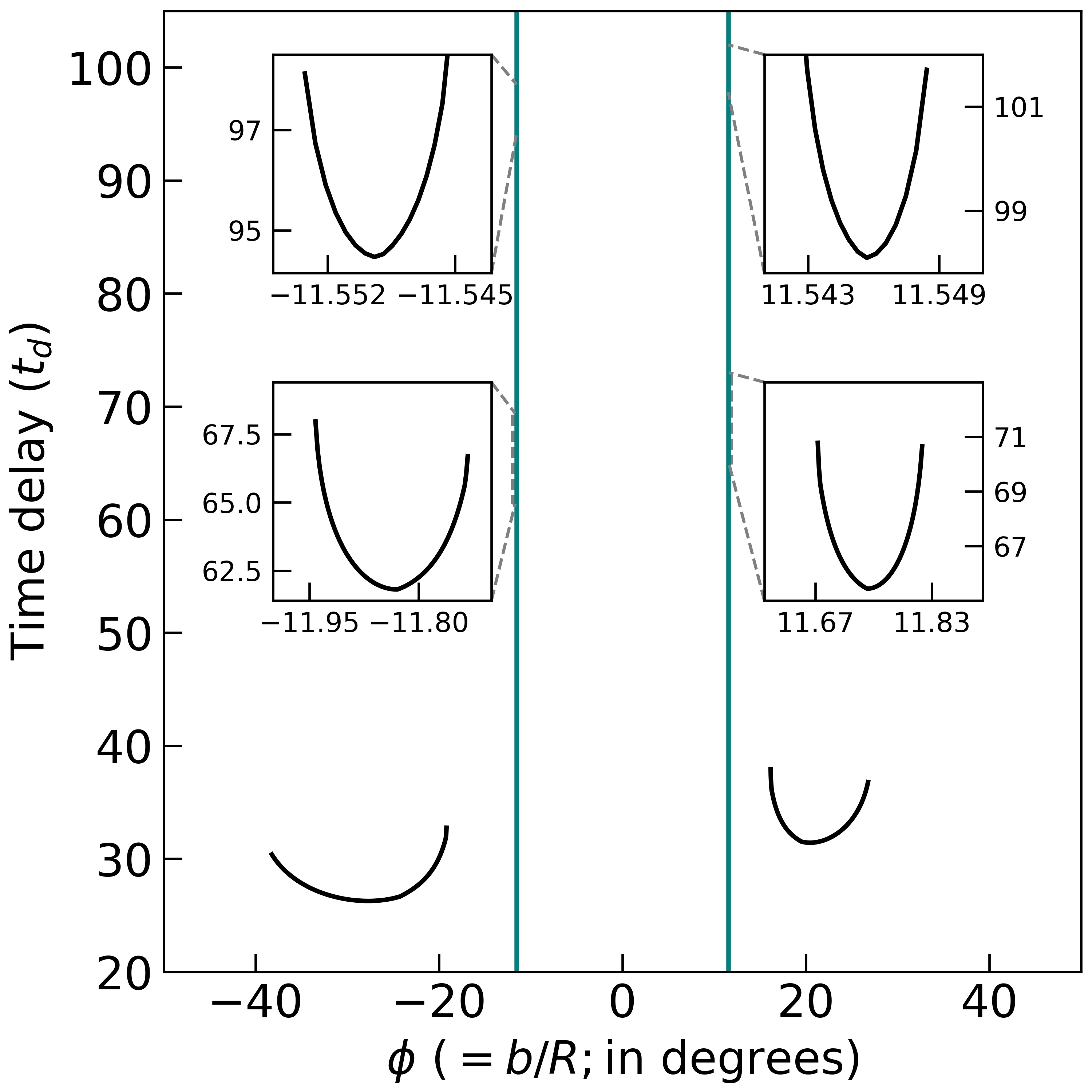}
    \caption{Wavefront propagation near a Schwarzschild black hole in
    the off-axis case. 
    \textit{Left panel:} The green, black and purple dots represent the 
    position of the source, lens, and observer, respectively. The green 
    curves represent the wavefront emitted by the source. The purple 
    and olive curves represent a wavefront emitted from the observer 
    at different time instances. The lensed images correspond
    to the points on purple/olive and green wavefronts where their normal
    vector agree with each other. These points are essentially the 
    points in the figure where purple/olive and green wavefronts touch 
    each other. The grey curves represent the corresponding rays 
    emitted from the source and reach to the observer.
    \textit{Right panel:} Time delay~($t_d$; in units of~$GM/c^3$) as 
    a function of angle of closest approach~($\phi$; in degrees) with 
    respect to the observer for off-axis case. The green vertical 
    lines mark the angle corresponding to the photon sphere. The black
    curves represent the slices of time delay surface near the primary, 
    secondary, and tertiary images as we go from small to large time 
    delays. For time delay surface near secondary 
    and tertiary images, we show zoom-in plots since the images 
    form very close to the photon sphere.}
    \label{fig:wavefront_OffAxis}
\end{figure*}

\begin{figure*}
    \centering
    \includegraphics[width=0.49\textwidth]{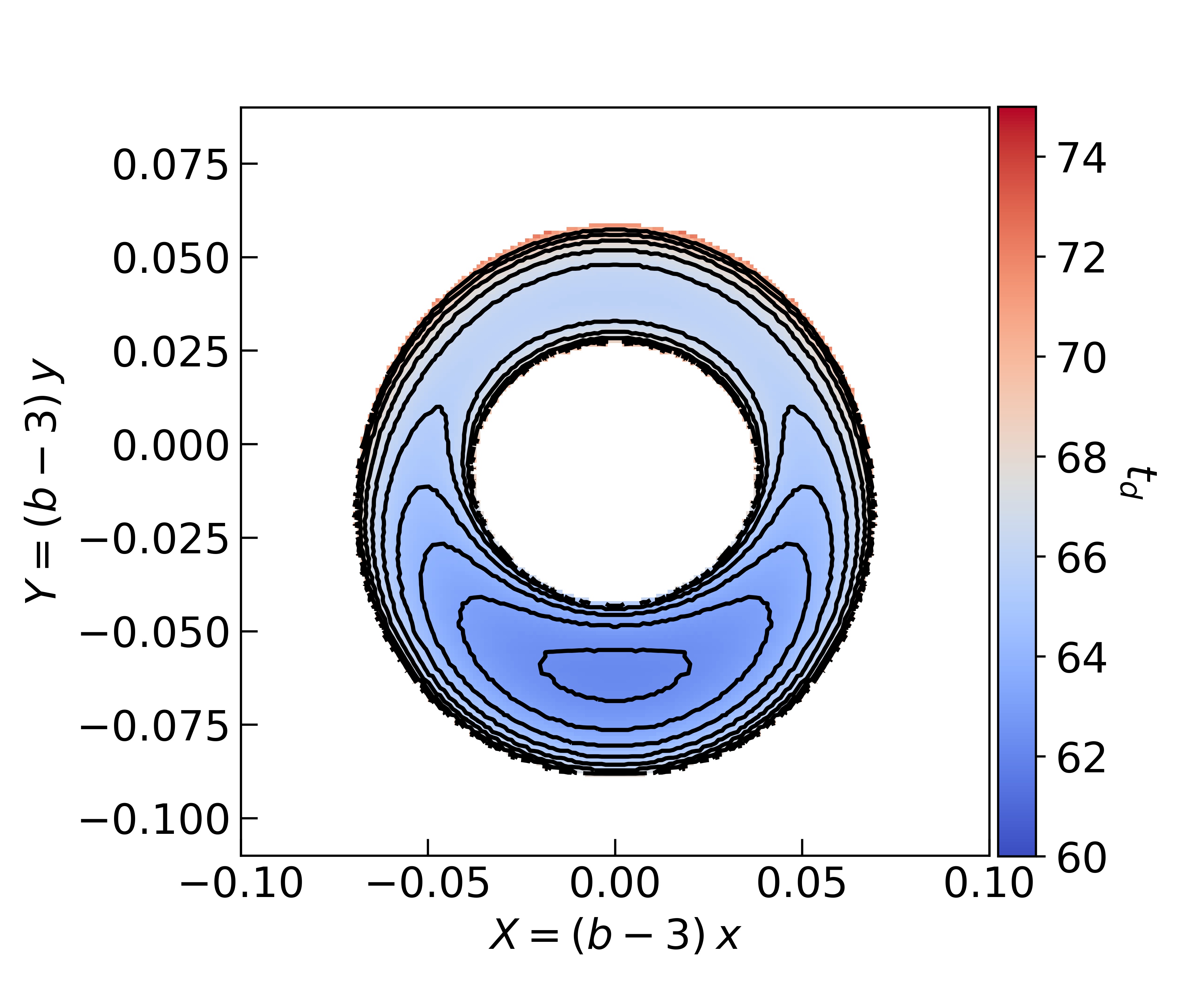}
    \includegraphics[width=0.45\textwidth]{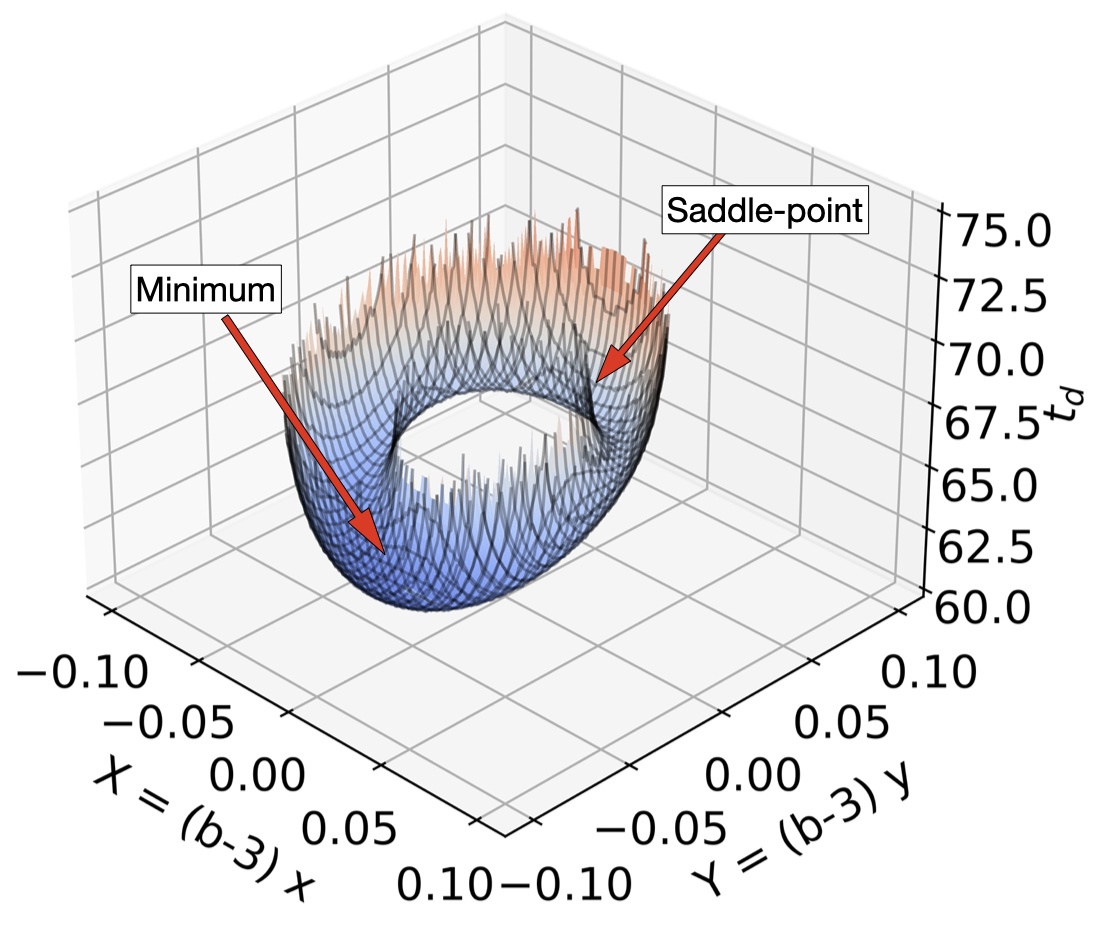}
    \includegraphics[width=0.49\textwidth]{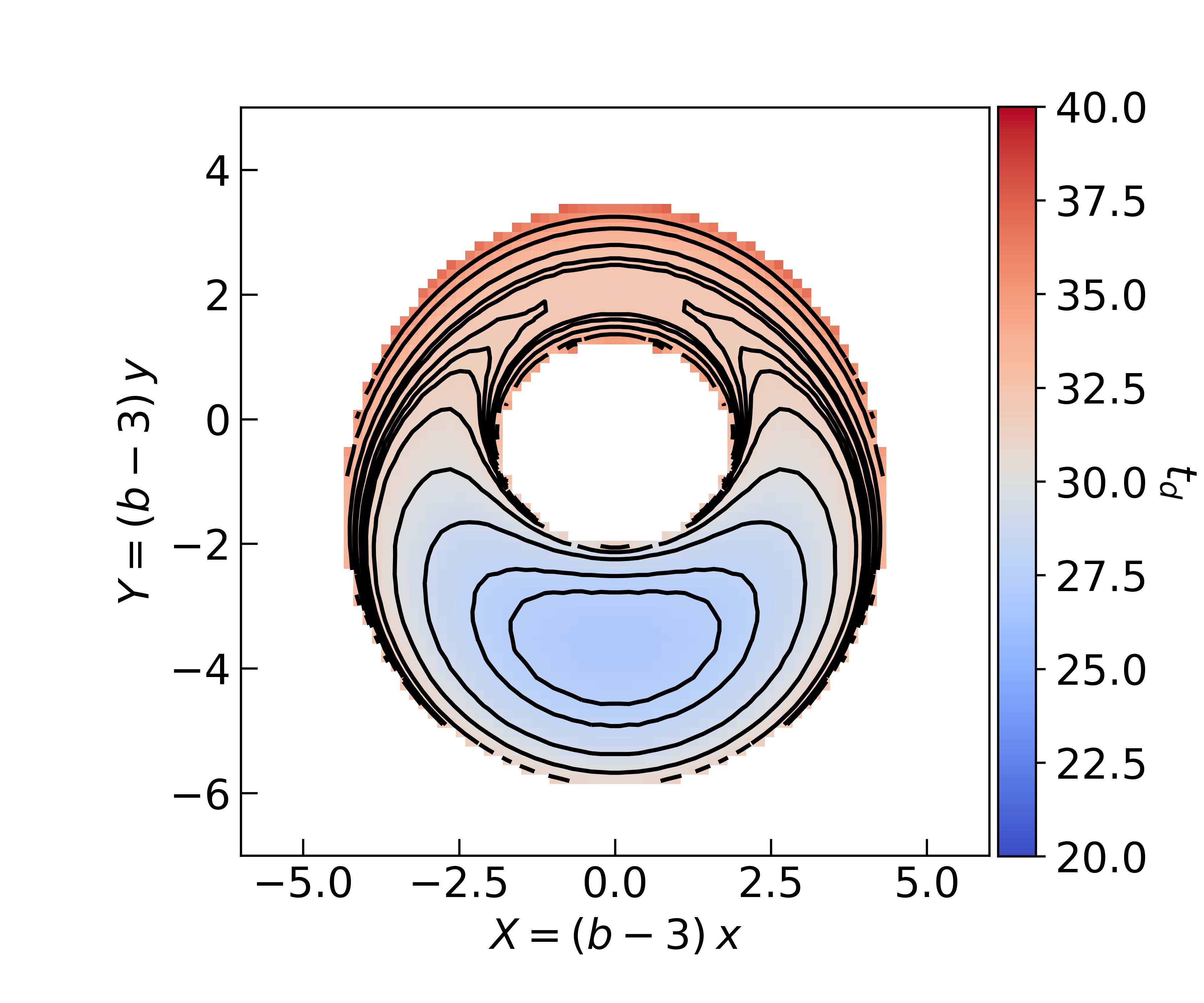}
    \includegraphics[width=0.45\textwidth]{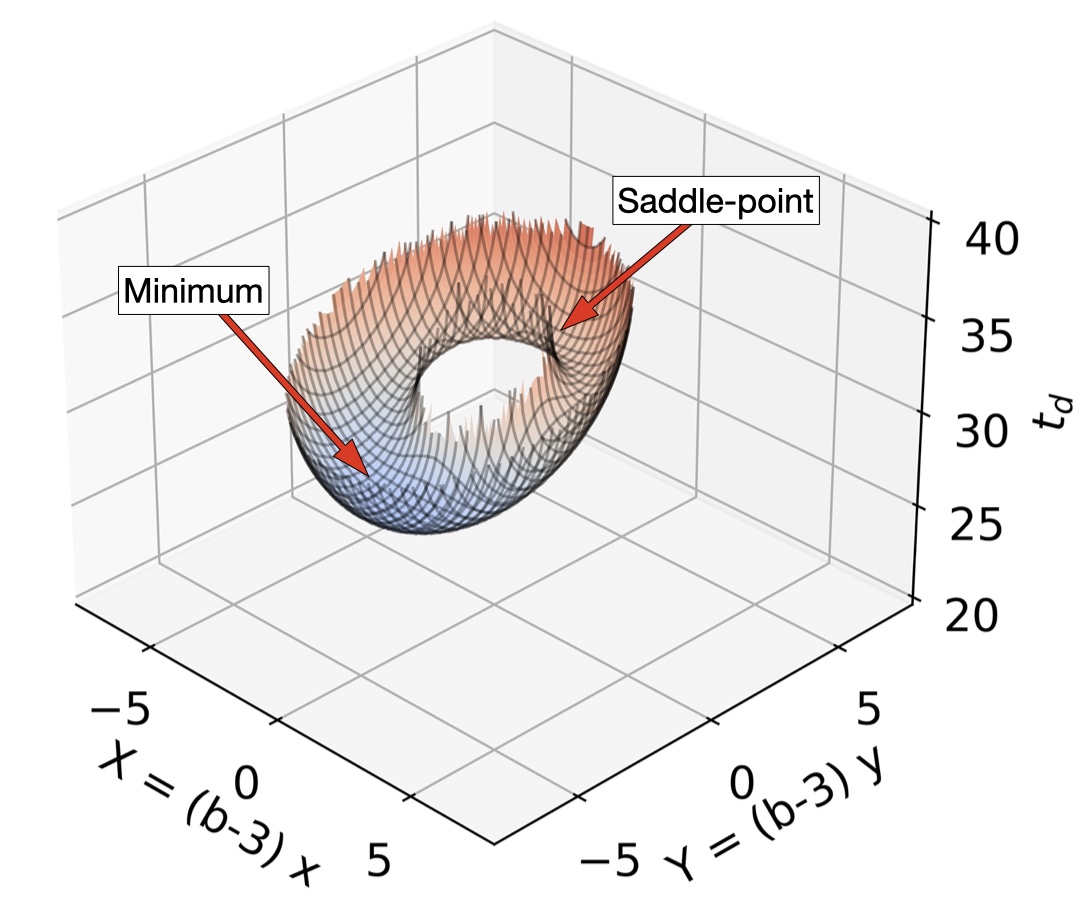}
    \caption{Time delay surface near primary and secondary image pairs 
    for the off-axis case. The left column represents the time delay 
    contours in 2D, whereas the right column represents the same plot 
    in 3D. In the right column, red arrows mark
    the positions of minima and saddle-points on the 3D time delay
    surface. The bottom and top rows correspond to the primary 
    and secondary image pairs respectively. In each panel, the
    X- and Y-axis are transformed to remove the region inside the photon
    sphere.}
    \label{fig:tds_OffAxis}
\end{figure*}

In our current work, to determine the path of light rays in the strong 
field and trace wavefronts originated from the source 
at~$(t, r,\phi)=(0,R,\pi)$, we solve Equation~\eqref{eq:eq_to_solve} 
numerically while choosing a range of~$b$ values.
To numerically solve the coupled differential Equation~\eqref{eq:eq_to_solve}, 
we use \texttt{odeint} from \texttt{SciPy}~\citep{2020SciPy-NMeth}.
An example of light rays and wavefront propagation near the 
Schwarzschild black hole is shown in Figure~\ref{fig:wavefront}.
In each panel, the black dot represents the black hole, and the dashed 
circle around it marks the photon sphere~($r=3M$).
The source is represented by the green dot and light rays emitted 
from the source are represented by the grey curves.
The corresponding equal time surfaces (or wavefronts) for three 
different time values~(in increasing order) are shown by green curves
in the three panels from left to right.
The break in the middle of the wavefront corresponds to the light
rays which fall inside the black hole~($b<3M$).
From Figure~\ref{fig:wavefront}, we can see that the rays passing 
closer to the black hole are deflected more strongly compared to far 
away rays.
Due to the large deflection close to the black hole, we observe part 
of the wavefront going around the black hole corresponding to the 
rays which loop around the black hole.

\section{Axially symmetric case}
\label{sec:axis_sym}

In this section, we consider wavefronts for an axially symmetric
configuration, i.e., the source, lens, and observer lie in a straight
line (the optical axis).

To construct the time delay surface in our current work, we use
forward and backward propagating wavefronts emitted by the source and 
observer, respectively, as described in~\citet{Nityananda_1990, 10.1007/BFb0009227}. 
Figure~\ref{fig:wavefront_touch} depicts the basic idea of using the 
wavefronts to locate the lensed image positions and construct the time 
delay surface.
The green, black, and purple dots mark the positions of the source, 
black hole, and observer, respectively.
We start by marking a forward propagating wavefront at a certain time 
as shown by the green curve. 
After that, we track a backward propagating wavefront emitted from 
the observer (shown by the purple curves at different times) and 
determine the time when it crosses the forward propagating wavefront.
When purple and green wavefronts touch each other such that their normal 
vectors agree with each other, they correspond to the path of light 
rays emitted from the source and observed by the observer.
This is further highlighted by the grey curves in
Figure~\ref{fig:wavefront_touch}.

To indicate the crossing points of the forward and backward 
propagating wavefronts, we trace the individual rays corresponding to 
the backward wavefront as shown by purple and grey curves in the left 
panel of Figure~\ref{fig:wavefront_td_axis}.
Grey (purple) curves mark the rays which~(do not) cross the forward 
propagating wavefront.
Whether a ray will cross the forward propagating wavefront or not will 
depends on the time at which we mark the forward propagating wavefront.
In addition, the exact number of times a ray crosses the forward 
wavefront will also depend on the temporal position of the forward 
wavefront.
If the forward wavefront is yet to cross the black hole, we can only 
get at most two crossings for a given ray.
However, once it crosses the black hole, we can have many
crossings~(in principle) since a ray can loop around the black hole 
many times.
We stop the forward wavefront before it crosses the black hole and 
only need the time corresponding to the first crossing for each ray.

For the axially symmetric case, time delay as a function of the
emission direction $\phi$ for a given ray is shown in the right panel
of Figure~\ref{fig:wavefront_td_axis}~(assuming~$M=1$).  The green
vertical lines mark the $\phi$-values corresponding to the photon
sphere~($r=3M$).  Any photon emitted at an angle smaller than this
will fall inside the black hole.  The black U-shaped curves show the
slices~(at~$y=0$) of a 3D time-delay surface near the primary,
secondary, and tertiary images as we go from small to large time delay
values. Since the secondary/tertiary images are formed when the light
rays do one/two loops around the black hole, they form very close to
the photon sphere, as can also be seen from the left panel. Due to the
axial symmetry, in this case, we will observe~(Einstein) ring
formations for the primary/secondary/tertiary image, and all of these
images are minima.
In Figure~\ref{fig:wavefront_td_axis}, we also see
that U-shaped curves for global minima seem to end 
around~$\pm40^{\circ}$ abruptly. Such a break is not physical and 
results from the fact that we do not have enough rays that can be 
used to determine the time delay surface at large angles. 
In principle, the curves will continue to go up (to infinity) as we
go to larger angles.

This ring formation and parity of images can be seen clearly in 
the 3D plot of the time delay surface as shown in 
Figure~\ref{fig:tds_axis}.
The $x$ and $y$~axes represent the spatial axes and $z$~axis shows the 
time delay values.
The spatial axes are re-scaled such that 
$(X, Y) = (b-3)\left(x, y\right)$ to omit the~$r<3M$ region.
The left, middle, and right panels show the time delay surface near
primary, secondary, and tertiary images, respectively.
In each panel, the ring formation is obvious, as are the 
corresponding types of images~(global minima for primary ring and 
local minima for secondary and tertiary rings).

From the right panel of Figure~\ref{fig:wavefront_td_axis}, we can 
see that the time delay surface near primary, secondary, and tertiary 
rings is not joined together, and there are gaps between them. 
This gap corresponds to rays that go behind the observer.
A few such rays (in purple) can be seen in the left panel of 
Figure~\ref{fig:wavefront_td_axis}.
Since the deflection will be continuous as we move closer to the black 
hole, there will always be a part of the forward wavefront between 
different order of images that will never reach the observer. 
We discuss this further in Section~\ref{sec:maxima}.

\section{Off-axis case}
\label{sec:off_axis}

The spherically symmetric nature of the Schwarzschild metric leads to
the formation of rings when the source lies on the optical axis.
However, once we move the source away from the optical axis, light
rays emitted from the source and travelling from one side of the black
hole will reach the observer earlier compared to the other side, and
the ring formation will break into two separate images.  
Breaking of the Einstein ring into two distinct images 
can also be understood from the fact that light rays travel in 
constant~$\theta$ planes around the Schwarzschild black hole due to its 
spherically symmetric nature. Hence, in the axis-symmetric case, there 
are infinite planes (for every~$\theta$ value) containing the source, 
black hole, and observer such that a light ray emitted from the source 
can be observed by the source in any of these planes. Once we move the 
source away from the optical axis, there is only one plane that contains 
the source, black hole, and observer. Hence, light rays emitted from 
the source and travelling only in this plane can reach the observer. 
Within this plane, there are only two paths that connect the source 
and observer, leaving us with two images of the source.
An example of
this is shown in Figure~\ref{fig:wavefront_OffAxis}.  Here, we move
the source~(shown by a green dot) to a negative $y$ value.  The
forward~(backward) propagating wavefront is shown in green~(purple)
color. We plot multiple temporal positions of the backward wavefront.
The olive wavefront also shows the backward propagating wavefront at a
larger time value than the purple colored wavefronts.  Since we moved the
source to the negative $y$~axis, the negative-$y$ part of the forward
wavefront~(lower half) will reach the observer first, implying that
image on negative~$\phi$ values will be observed first by the
observer.  This can also be seen from the fact that the lower half part of
the green wavefront touches the last purple wavefront, whereas the upper
half of the green wavefront touches the olive wavefront~(which is drawn
for a larger time).  The light ray paths corresponding to the primary
lensed images are shown by the grey curves.  The breaking of ring in
two different images can be more clearly seen in right panel of
Figure~\ref{fig:wavefront_OffAxis} where we again plot time
delay~($t_d$) as a function of angle of closest approach~($\phi$) for
different rays as we observe that images formed on~$\phi<0$ has
smaller time delay values compared to the corresponding counterparts
on~$\phi>0$. Similar to Figure~\ref{fig:tds_axis}, 
we observe that the U-shaped curves corresponding to primary images
abruptly end near~$\pm40^{\circ}$. Again, this is not physical and is a 
result of the fact that we do not have enough rays to draw the curves at these 
angles; otherwise, they would continue to go up to infinity.
Another obvious yet important observation is the fact that in each 
(primary/secondary/tertiary) pair of lensed images, the image arriving 
later forms closer to the black hole.

Here, we can again ask for the parity of each of these images, but
the~$\phi-t_d$ plot shown in the left panel is not sufficient 
to determine the parity of these images since it only shows a 1D slice
along $y$-axis of the full 3D time delay surface.
Hence, we construct 2D as well as 3D time delay surface plots near 
the primary and secondary images, as shown in 
Figure~\ref{fig:tds_OffAxis}.
The 2D contour plots for primary and secondary images are shown in the 
bottom and top panel of the left column, respectively.
The corresponding 3D plots are shown in the right column.
From the left and right columns, we can clearly observe that the 
primary as well as secondary pair of images consist of one minima and one 
saddle-point. That said, it can be hard to locate the 
position of minima and saddle-points in the 3D plots~(right panel), so
we have explicitly pointed out their positions by red-coloured arrows.
Due to the spherical symmetry of the lens, we expect to see the same 
image types even for higher-order images.
To plot the time delay surface, we again use the same change of axes, 
$(X, Y) = (b-3)\left(x, y\right)$, and omit the~$r<3M$ region.
Similar to the axis-symmetric case, we again observe a gap in the time 
delay surface between primary and secondary images.

\section{The ``Home'' and ``Away'' Images}
\label{sec:maxima}

In both of the above cases~(axially symmetric and off-axis), we
observed gaps in the time delay surface between each order of lensed
images as seen from left panels in Figure~\ref{fig:wavefront_td_axis}
and~\ref{fig:wavefront_OffAxis}.  As mentioned briefly in
Section~\ref{sec:axis_sym}, these gaps correspond to the part of the
backward propagating wavefront that goes behind the observer and never
crosses the forward wavefront.  Or, from the forward wavefront
perspective, part of the wavefront that loops around the black hole
and goes behind the source itself.  Since the deflection angle is
continuously increasing as we move closer to the black hole, there
will always be a part of the forward~(backwards) propagating wavefront
that will go behind the source~(observer).  We remark that in standard
lensing theory, singular lenses such as a point mass, which do not
explicitly invoke black holes, also have similar gaps in time delay
surface, which can be considered as the infinitely time delayed maxima
forming at the position of the point
lens~\citep[e.g.,][]{2002glml.book.....M} assuming that the time delay
surface is continuous.

\begin{figure}
    \centering
    \includegraphics[width=0.46\textwidth]{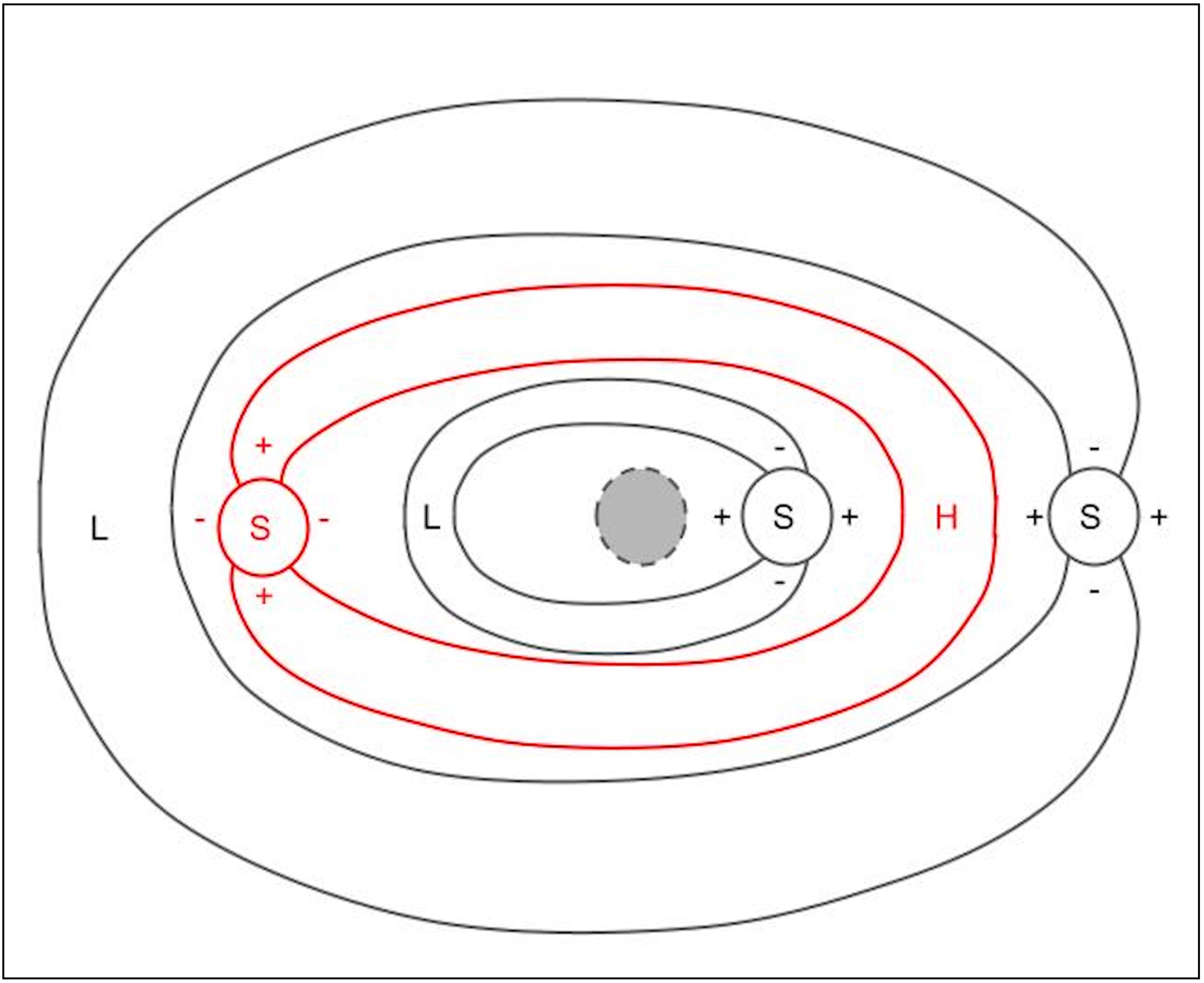}
    \caption{A schematic representation of the time-delay surface topology
    near a Schwarzschild black hole for the off-axis case\citep[extending
    Figures 5.7 and 5.8 in ref.][]{1992grle.book.....S}.
    The position of the 
    black hole is shown by the gray shaded region encircled by dashed 
    curve at the centre of the plot. ``L'', ``S'', and ``H'' denote 
    the position of minima, saddle-point, and maxima, respectively.
    The `+' and `-' signs around saddle-points indicate the directions
    in which time delay values are increasing and decreasing compared
    to time delay values at the saddle-points, respectively.  The
    image positions in black, marks the position of observed images
    (or ``home'' images) where image positions in red mark the 
    infinitely delayed images~(or ``away'' images) which we expect to 
    form in gaps between the home images. Here we only show the first 
    two orders of home images and one set of away images, but the same 
    topology is expected even for the higher-order images.}
    \label{fig:contours}
\end{figure}

Near a Schwarzschild black hole, we can again ask the question,
assuming that the time delay surface is continuous, do we expect
additional images to form in these gaps between observed images?  To
answer this question, we need to determine the time delay surface
topography near the black hole.  A schematic surface is shown in
Figure~\ref{fig:contours} for the off-axis case.  The position of the
black hole is shown by the grey shaded region inside the dashed
circle.  Here, the type of images are marked by ``L'', ``S'', and
``H'' for minima (or low), saddle-point, and maxima (or high),
respectively.  The black markers show what we may call ``Home'' image
positions, i.e., images observed by the observer.  On the left side of
the black hole, we see the formation of two such minima, whereas on the
right side, we see the formation of two such saddle-points.  These
correspond to the first two orders of images~(primary and secondary)
shown in Figure~\ref{fig:tds_OffAxis}.  In between these home minima
(saddle-points), we show the formation of an infinitely delayed
saddle-point~(maxima) shown in red, which we call ``Away'' images since
they never reach the observer.  Although here we only show the
topography near primary and secondary home images, we expect
the same for further higher-order images due to the spherical symmetry
of the lens.  Since the proposed images in gaps are not observed, 
the above topology makes the time delay surface continuous.  In
addition, in the above topology, one (global) maxima will also form at
the position of the black hole~($r<3M$).  Doing so, in addition to
making the time delay surface continuous, also satisfies the odd image
theorem~\citep{1981ApJ...244L...1B}.  In addition, if we 
move the source also on the observer side (or observer 
on the source side) of the black hole by changing the sign of the
x-coordinate, then the earlier home images become away images and 
earlier away images become home images.

\section{Conclusions}
\label{sec:conclusions}

The formation of multiple images through gravitational lensing is now
a commonplace in astronomy and has a well-developed theoretical
formalism to interpret the observables.  This formalism assumes weak
gravitational fields, and consequently small deflections, which do not
apply to the multiple images formed near a black hole, such as
observed near the M87 black hole.  Can the existing formalism be
generalised to these strong-field applications?

In this work we show that the key element of the weak-field formalism
does generalise rather simply.  This element is the abstract
construction known variously as the time-delay surface, the
arrival-time surface, or the Fermat potential.  Lensed images form at
the zero-gradient locations of the surface (maxima, minima, and
saddle-points), and higher derivatives give various properties of
images, such as the apparent handedness or parity.  In the weak-field
formalism, the time delay surface is conventionally given by the sum
of two contributions, one geometrical and one gravitational, to the
light travel time.  In strong fields, it is not clear how to identify
two such separate contributions.  However, an alternative definition
of the time-delay surface, as the difference between a forward and a
backward wavefront, can be applied to any static spacetime.  We
compute the time-delay surface near a Schwarzschild black hole and
study its properties.

Concretely, we use crossings of forward and backward propagating
wavefronts from the source and observer, respectively, to construct the
time delay surface and determine the image types.  In the axially
symmetric case (having the source, lens centre, and observer on the same
line), we observe ring formation corresponding to a minimal valley on
time delay surface for primary~(1st order), secondary~(2nd order), and
tertiary~(3rd order) images.  Moving the source away from the optical
axis causes the ring to break into two separate images, one minimum and
one saddle-point.  We again show this by constructing the time delay
surface near the primary as well as secondary images.  The pattern
will continue since near a Schwarzschild black hole there is an
infinite sequence of images with continuously decreasing magnification
factors (see Appendix~\ref{sec:app3}) as we move towards higher order
in the sequence.

In between each ring (or pair of lensed images), we find 
steeply rising walls in the time delay surface.  
These walls are a
result of the fact that near the black hole light rays can loop around
the black hole and between each order of images there will be a
certain fraction of rays emitted from the source that will never reach
the observer and go behind the source itself.
Between a pair of walls we can think of two images,
one saddle and one maximum, both infinitely
delayed and therefore not visible.  We name these \textit{away}
images, as distinct from the observable \textit{home} images.  A final
image, infinitely demagnified and infinitely delayed, will form within
the photon sphere~($r<3M$).  The odd-image theorem remains valid.  To
an observer on the same side of the black hole as the source, \textit{home} and
\textit{away} images get swapped, as do minima and maxima.

This work has been limited to a Schwarzschild black hole, for which we
have mainly offered heuristic arguments from examining the numerical
results on the time-delay surface.  Formulating the image properties
more precisely in terms of the surface is desirable, but it is not
obvious how to proceed.  The time-delay surface for a Kerr black hole,
which would be more representative of the observations, would be
interesting to compute, though significantly more complicated than the
Schwarzschild case.

\section*{Acknowledgements}
The authors thank Jasjeet Singh Bagla, Rajaram Nityananda, Dominique Sluse, 
and Liliya Williams, and the anonymous referee for useful comments.
A.K.M. acknowledges support by grant 2020750 from the United 
States-Israel Binational Science Foundation (BSF) and 
grant~2109066 from the United States National Science 
Foundation~(NSF), and by the Ministry of Science \& Technology, 
Israel. 
This research has made use of NASA's Astrophysics Data System 
Bibliographic Services.

The work utilises the following software packages:
\textsc{Python}~(\url{https://www.python.org/})
\textsc{NumPy}~\citep{2020Natur.585..357H}, 
\textsc{Matplotlib}~\citep{2007CSE.....9...90H},
\textsc{SciPy}~\citep{2020SciPy-NMeth}.

\bibliographystyle{apsrev4-2}
\bibliography{Reference}  

\appendix
\section{Null geodesics in Schwarzschild spacetime}
\label{sec:app1}

Lensing by a Schwarzschild black hole is a classical topic in the
literature and there are several ways of computing light paths.  One
elegant approach is to treat $\frac12 g^{\mu\nu} p_\mu p_\nu$ as a
Hamiltonian~$(\mathcal{H})$ in four dimensions with canonical
momentum $p_\mu\left(=\partial\mathcal{H} / \partial \dot \mu\right)$
being a new abstract vector and the affine parameter as the
independent variable~\citep[e.g.,][]{1983mtbh.book.....C}.

The Schwarzschild metric itself is an exact static, spherically
symmetric solution of Einstein's field equations in
vacuum\citep{1916SPAW.......189S}.  Because of spherical symmetry,
geodesics will be confined to a plane, and hence it is sufficient to
consider the spacetime slice $(t, r, \phi)$.  Omitting factors
of $G$ and $c$ the metric can be written as
\begin{equation}
    ds^2 = -\left(1-\frac{2M}{r}\right) dt^2 
           + \left(1-\frac{2M}{r}\right)^{-1} dr^2 
           + r^2 d\theta^2
           + r^2 \sin^2\theta \: d\phi^2,    
    \label{eq:sch_metric}
\end{equation}    
where $M$ is the mass of the black hole.  This metric leads us to
\begin{equation}    
    2\mathcal{H} = -\left(1-\frac{2M}{r}\right)^{-1} p_t^2 + 
    \left(1-\frac{2M}{r}\right) p_r^2 + {p_\phi^2 \over r^2},
    \label{eq:hamiltonian}
\end{equation}
with $\mathcal{H}=0$ for null geodesics~(i.e., light rays).  
Solving Hamilton's equations give
\begin{equation}
    \begin{aligned}
        \dot t &= - \left(1-\frac{2M}{r}\right)^{-1} p_t, \\
        \dot r &= \left(1-\frac{2M}{r}\right) p_r, \\
        \dot \phi &= {p_\phi \over r^2}.
    \end{aligned}
    \label{eq:Heq}
\end{equation}
We can derive the constants of motion  using the equations,
\begin{equation}
    \begin{aligned}
        \dot p_t &= 0 \quad \Rightarrow \quad \left(1-\frac{2M}{r}\right) \dot t = E \: ({\rm constant}), \\ 
        \dot p_\phi &= 0 \quad \Rightarrow \quad r^2 \dot \phi = L \: ({\rm constant}).
    \end{aligned}
    \label{eq:Heq_zeros}
\end{equation}
For light rays,~$E$ and~$L$ are equivalent to the energy 
and angular momentum of the ray.
The constant value of $p_t\:(=E)$ is arbitrary, and we can
choose it to be -1 by rescaling the affine parameter. 
There is a non-trivial equation for $p_r$, but 
since $\mathcal{H}=0$ for light rays, we can simply
eliminate $p_r$ to get
\begin{equation}
\dot r = \sqrt{1-\frac{p_\phi^2}{r^2}\left(1-\frac{2M}{r}\right)}.
\label{eq:rdot}
\end{equation}
The square root implies $r\geq b$, where $r=b$ represents the closest
approach, which is defined by
\begin{equation} 
    p_\phi^2 = {b^3\over b-2M}.
    \label{eq:rmin}
\end{equation}
Putting $r=b$ is equivalent to setting~$\dot r=0$ and~$p_r=0$, which
are the conditions that a circular orbit will have and leads
to~$r=3M$~(i.e., a photon sphere).  With the above, Hamilton's
equations~(\ref{eq:Heq}) now reduce to
\begin{equation}
    \begin{aligned}
    {dr \over dt} &= \left(1-\frac{2M}{r}\right)  \dot r, \\
    {d\phi \over dt} &= \left(1-\frac{2M}{r}\right) {p_\phi \over r^2},
    \label{eq:eq_tosolve}
    \end{aligned}
\end{equation}
determining the trajectory of rays emitted by the source and passing
near the black hole. Eq.~(\ref{eq:eq_to_solve}) in the main text is
simply Eqs.~(\ref{eq:rdot}--\ref{eq:eq_tosolve}) collected together.

The differential equations (\ref{eq:eq_tosolve}) can be easily
integrated numerically, starting from $r=b$ and any initial $\phi$.
Some care is needed, however, because $dr/dt$ is not well-behaved at
$r=b$.  We can change variable to $w$, where
\begin{equation}
r = b + w^2 \qquad
\end{equation}
which transforms the $r$ equation to
\begin{equation}
  \frac{dw}{dt} = \left(1-\frac{M}{b+w^2}\right) \frac{\dot r}{2w}
\end{equation}
which is well-behaved at $w=0$ (that is, $r=b$) because
\begin{equation}
\frac{\dot r^2}{(2w)^2} = \frac{b-3}{2(b-2)} + O(w^2)
\end{equation}
This last quantity, it turns out, can actually be expressed as
\begin{equation}
\sum_{n=0}^\infty \left(\frac{-1}b\right)^{n+1}
\frac{(n+2)(n+3-b)}{4(b-2)} \; v^{2n}
\end{equation}
but only the $w\rightarrow 0$ limit is needed.

\section{Limiting cases}
\label{sec:app2}
The above equations have three interesting limiting 
cases for different~$b$ values,
\begin{enumerate}
    \item $b\gg M$: This case corresponds to the
    weak field approximation. Even in the weak field 
    regime, we have conventional terms like, ``strong 
    lensing'' and ``weak lensing''. In weak field limit, 
    strong lensing refers to the case where we observe
    formation of multiple images.
    \item $b \ll R$:  we have a small lens, which 
    behaves as a deflector of straight light paths.
    \item $b\rightarrow 3M$: A photon does multiple 
    orbits around the black hole before going away.
    If $b<3M$ the photon falls into the black hole.
\end{enumerate}

The standard astrophysical scenario satisfies the
first two limits, i.e.,~$R\gg b \gg M$.
In such a case, to determine the total deflection
angle, it is useful to write~$\phi$ as a function
of~$r$,
\begin{equation}
    {d\phi \over dr} = {1 \over r^2}
    {1 \over \sqrt{{b-2M \over b^3} - {r^3 \over r-2M}}}.
    \label{eq:eq_weak}
\end{equation}
The total deflection angle experienced by a light ray
emitted from a source at~$(r,\phi)=(R,\pi)$ is
\begin{equation}
    \phi=\pi - 2\int_b^R {dr \over r^2 \sqrt{{b-2M \over b^3} - {r^3 \over r-2M}}}.
    \label{eq:eq_weak_int}
\end{equation}
Here, we introduce a change of 
variable,~$u={R-r \over R-b}{b \over r}$ such that
\begin{equation}
    \phi = \pi - 2 \int_0^1 
    {du \over \sqrt{1-u^2-\frac{2M}{b}(1-u^3)+\frac{2b}{R}(1-u)}}.
    \label{eq:eq_weak_int_u}
\end{equation}
Binomial expansion to first order in~$M/b$ and~$b/R$ 
leads to
\begin{equation}   
    \phi = \pi - 2 \int_0^1 {1 \over \sqrt{1-u^2}}
    \left( 1 + {1+u+u^2 \over 1+u} {M \over b} \right. 
               \left. - {1 \over 1+u} {b \over R}  \right).
\end{equation}
Solving the above integral after substituting~$u=\cos w$
leads to,
\begin{equation}
    \phi \simeq 2\frac{b}{R} - \frac{4M}{b},
    \label{eq:weak_le}
\end{equation}
which is essentially a straight line with 
an extra deflection of~$4M/b$.

On the other hand, in the strong field 
limit~($b\rightarrow3M$) with~$R\rightarrow\infty$, 
we replace~$b$ by~$3M(1+\epsilon)$ and introduce a 
change of variable such that
\begin{equation}
    r = 3M \frac{1+\epsilon}{1-v}.
\end{equation}
Since, in such a case, the integral in Equation~\eqref{eq:eq_weak_int}
will be dominated by contribution near~$v=0$, we
have
\begin{equation}
    \phi \simeq \pi + 2 \int_0^1 \frac{dv}{\sqrt{v^2+\frac{2}{3}\epsilon}},
\end{equation}
leading to
\begin{equation}
    \phi \simeq 2 \ln(\epsilon) + \dots
\label{eq:logeps}
\end{equation}

\section{Magnification}
\label{sec:app3}

To calculate magnification, we need to set up a correspondence between
lensed and unlensed rays.  One could argue for different ways of doing
so, but one reasonable definition for an unlensed ray is to require it
to have the same value of $p_\phi$ as the lensed ray.  The unlensed
ray travels in a Euclidean line to some $\bar\phi$ (say).  From the
geometry it is easy to see that (i)~the unlensed ray makes an angle
$\half\bar\phi$ with the $x$ axis, and (ii)~the closest approach of
this ray to the origin is $R\sin\half\bar\phi$.  Since for $M=0$ the
closest approach is simply $p_\phi$ we have
\begin{equation}
    \sin\half\bar\phi = \frac{p_\phi}R
\label{def-barphi}
\end{equation}
In other words, the unlensed ray is a line from $(R,\pi)$ to
$(R,R\bar\phi)$, where $\bar\phi$ is given by (\ref{def-barphi}).
The derivative $\pderiv(\bar\phi/\phi)$ is a possible definition of
magnification in the $\phi$ direction.

More interesting, however, is the ratio of lensed and unlensed solid
angles, since it corresponds to light flux.  Going to three
dimensions, and considering the solid angles within $\phi$ and
$\bar\phi$ we have
\begin{equation}
    \Omega = 4\pi \, \sin^2 \! \half\phi \qquad
    \bar\Omega = 4\pi\, \frac{p_\phi^2}{R^2}
\end{equation}
The solid-angle magnification (sometimes called amplification) is
\begin{equation}
     A \equiv\pderiv(\bar\Omega/\Omega)
     = \frac4{R^2} \, \frac{p_\phi}{\sin\phi} \, \tderiv(p_\phi/\phi)
\label{scalmag}
\end{equation}
The amplification becomes singular at $\phi=0$.  The regime of
$|\bar\phi|$ small and $|\phi|$ large corresponds to strong-field
lensing.  The curve is nearly flat, indicating very faint images, but
there are tiny step-like features at multiples of $180^\circ$.  The
regime $|\phi|$ large and $\bar\phi\simeq\phi$ is where the light is
always far from the lens.  Here the slope is close to unity --- but
must be slightly less than unity to compensate for the steep regions.

Let us now consider limiting forms of the amplification~(\ref{scalmag}).

For the most-deflected part of the wavefront, we take the
$b\rightarrow3M$ limit, corresponding to Eq.~(\ref{eq:logeps}).  This
gives
\begin{equation}
     A = \frac\epsilon{R^2\sin\phi} \; \tderiv(p_\phi^2/\epsilon)
         \propto {e^{-|\phi|} \over \sin\phi}
\end{equation}
Thus, the flux in the later images falls off quickly.

For small angles, we have
\begin{equation}
     A = \frac{\bar\phi}\phi \; \tderiv(\bar\phi/\phi)
\end{equation}
From the lens equation (\ref{eq:weak_le}) we have
\begin{equation}
    \phi = \bar\phi - (2\phie)/\bar\phi
\label{lenseq-phi}
\end{equation}
where we have defined
\begin{equation}
    \phie \equiv \frac{2M}R
\end{equation}
the conventionally definition of the Einstein radius for the case of
observer-lens and lens-source distances both equal to $R$.
For the flux we get
\begin{equation}
     A = \left(1-(2\phie/\bar\phi)^4\right)^{-1}
\end{equation}
Now, from the form of (\ref{lenseq-phi}) it is clear that there will
be two values of $\bar\phi$, one each greater and less in magnitude
than $2\phie$.  Hence, for small angles, there is always one image
with amplification greater than unity.  This is the flux-conservation
paradox.  Its resolution depends on precisely how the amplification is
defined --- recall that the definition (\ref{def-barphi}) is not
unique --- but basically, the answer is that at large angles the
amplification dips very slightly below
unity\cite{2008MNRAS.386..230W}.

\end{document}